\documentclass[conference]{IEEEtran}
\IEEEoverridecommandlockouts
\usepackage{cite}
\usepackage{amsmath,amssymb,amsfonts}
\usepackage{algorithmic}
\usepackage{graphicx}
\usepackage{textcomp}
\usepackage{xcolor,float}
\usepackage{authblk}
\usepackage{xspace}
\usepackage{todonotes,comment}

\usepackage[hang,small,bf]{caption}    

\usepackage{tikz}	
\usetikzlibrary{backgrounds,fit,decorations.pathreplacing}  

\def\BibTeX{{\rm B\kern-.05em{\sc i\kern-.025em b}\kern-.08em
    T\kern-.1667em\lower.7ex\hbox{E}\kern-.125emX}}
\begin{document}

\title{Solving  machine learning  optimization problems using quantum computers\\
}

\author[a]{Venkat R. Dasari, Ph.D.\thanks{Send correspondence to VRD:  venkateswara.r.dasari.civ@mail.mil}}
\author[b]{Mee Seong Im, Ph.D.\thanks{Mee Seong Im: meeseong.im@westpoint.edu}}
\author[c]{Lubjana Beshaj, Ph.D.\thanks{Lubjana Beshaj: lubjana.beshaj@westpoint.edu}}
\affil[a]{U.S. Army Research Laboratory, Aberdeen Proving Ground, MD 21005}
\affil[b]{U.S. Military Academy, West Point, NY 10996}
\affil[c]{Army Cyber Institute, West Point, NY 10996}

\maketitle

\begin{abstract}
Classical optimization algorithms in machine learning often take a long time to compute  when applied to a multi-dimensional problem and require a huge amount of CPU and GPU resource. Quantum parallelism has a potential to speed up machine learning algorithms. We describe a generic mathematical model to leverage quantum parallelism to speed-up machine learning algorithms. 
We also apply quantum machine learning and quantum parallelism applied to a $3$-dimensional image that vary with time. 
\end{abstract}


\section{Introduction}

Machine learning (ML) is generally considered an optimization problem and  sometimes they are computationally intense, requiring HPC computing resource to complete the optimization. Classical  algorithms do not do well when dealing with high dimensional problem space. Optimization problems in ML often  reach NP-hard in their complexity  and will be difficult to handle by the classical computers. One of the benefits of quantum computing is its parallelism which can speed up the  optimization computations to  quickly compute global minimum and maximum. In our previous work we have described the programmable quantum networks to support quantum applications using a unified quantum channel, achieved through openflow abstractions to carry quantum metadata on classical channels\cite{dasari2019openflow}. In a different study \cite{im2018optimization}, we have also described a mathematical model to synchronize the quantum and classical channels to coordinate the performance of quantum application on a multi-node quantum network topology.

\begin{figure}[!h]
	\includegraphics[width=\linewidth]{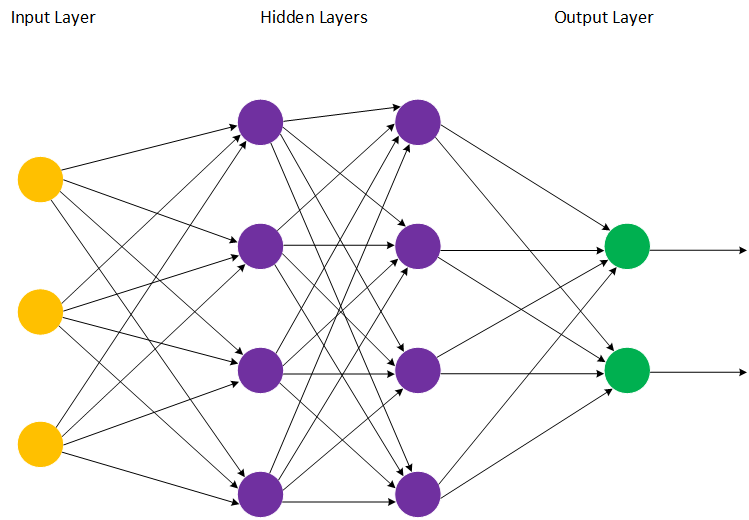}
	\caption{Neural network  with one input layer, two hidden layers and one output layer}
	\label{fig:boat1}
\end{figure}


\section{Machine learning}

Machine learning is a branch of Artificial Intelligence (AI), where the   generalized mathematical analytical model is build from data  for analysis and pattern recognition. 

The machine is  learning an optimization problem and the complexity increases with the size and dimensionality of the input data. Since quantum computers are well designed to handle tensor and dot products, they are highly suited for machine problems involving complex, high dimensional data in order to speed up the learning process. In a complex data set, feature selection involves an exhaustive search, whose complexity often reaches exponential, making it impractical to undertake any such effort. The users end up using heuristic methods with polynomial-time complexity. However, quantum computers can accomplish such optimizations with polynomial-time complexity, using parallel computation and without having a need for imperfect heuristic optimization methods. In machine learning, most data is commonly represented as $2$-dimensional arrays and input layers are mostly in the form of $2$-dimensional arrays, but for image processing, the input layer can be $3$-dimensional to accommodate the color channel, in addition to height and width of the pixel.

Machine learning and deep learning  use a hypothesis function  to formulate the machine learning problem using theta parameters as shown in 
\begin{equation}\label{eq-1}
	h_{\theta}(x)=\theta_{0}+\theta_{1}x_{1} +\ldots+\theta_{n}x_{n}, 
\end{equation} 
where $\theta=(\theta_0,\theta_1, \ldots, \theta_n)$ and 
$x=(x_1,\ldots, x_n)$. 
 Each $x_i$-value represents the features that allow the mathematical model to construct a computational graph. That is, we need to find best parameters to create a model, and train the model using training data to make accurate predictions. Once the hypothesis function is constructed using parameters, features and biases, we need to build regression  and loss functions to optimize the neural network models.

We compute the cost function by summing over the square of the difference of the observed and the expected values for the entire set of $\theta$ parameters to arrive at a cost or loss values for each training data set ran against the model: 
\begin{equation}
\displaystyle \min_{\theta_0,\theta_1}\frac{1}{2m}\sum_{i=1}^{m} \left(   h_{\theta}(x^{(i)})     - y^{(i)} \right)^2. 
\end{equation}

Computational complexity of the del learning algorithms increases in relation to the size of input space and  number of the parameters needed to be optimized for the model. They also need  large quantities of training data to  build a mathematical model that predicts accurately. All the tasks related to model building and training  can take  leverage of the parallelism offered by quantum computers.

\begin{figure}
	\includegraphics[width=\linewidth]{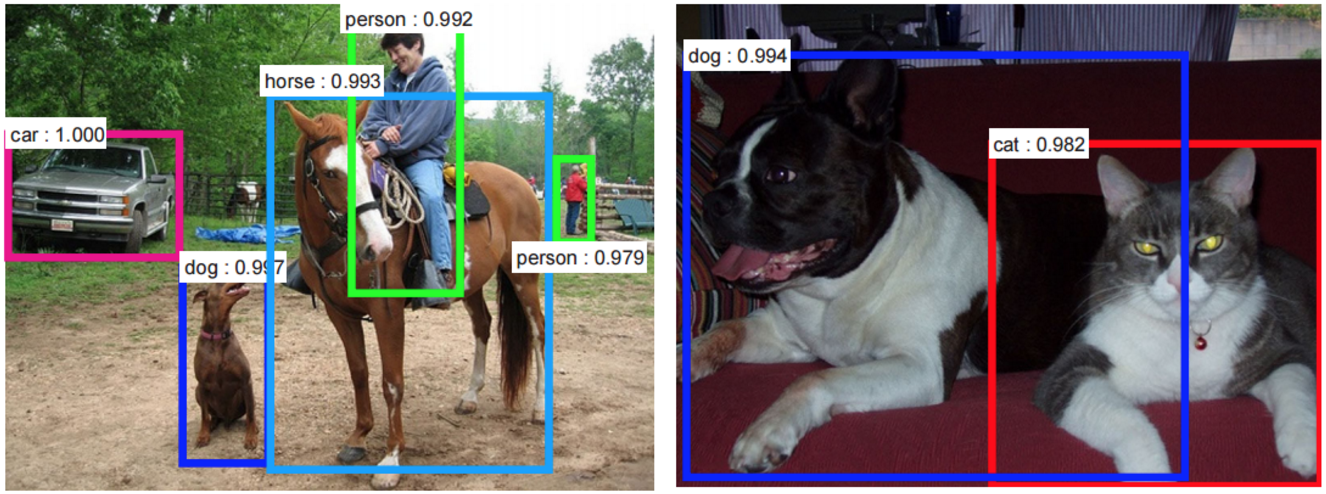}
	\caption{Example of CNN for object detection in live video streams. Source: https://ujjwalkarn.me/2016/08/11/intuitive-explanation-convnets/}
	\label{fig:boat1}
\end{figure}

As the  size of input space increases and with more hidden layers of complex multi-dimensional feature map, an optimization of the computational complexity of deep learning frameworks increases  tremendously. Advanced machine learning techniques use complex multi-layered convolutional neural networks (CNN) for image recognition in building complex scenarios. CNNs feature maps can be  a high dimensional cluster and  are computationally complex to execute\cite{kim2014convolutional}.

Long Short-Term Memory (LSTM) unit, Gated Recurrent Unit (GRU), and  
Recurrent Neural Networks (RNN)   
are among the most widely used models in deep learning for complex pattern recognition\cite{graves2013hybrid}. Training these models require intense computations and  often take very long time to compute  by classical computers. Quantum parallelism can reduce the computational time if deployed for training these models.

This is the reason why graphics processing unit (GPU) computing resources are deployed to compute  CNNs to speed up the computational process. Still  the complexity of feature maps can reach a point that even the addition of GPU resources will not compute the CNN in polynomial time, and such cases are good candidates for the  quantum computers, which can reduce the  computational complexity by quantum parallelism.

\section{Quantum computation}

Quantum mechanical principles are exploited to create quantum computers. A qubit is  a fundamental unit of quantum computation and  it exists in a binary state. Any system that exhibits binary states like  polarized photons can be used to create qubits. Qubit states of  0 and 1 are used as basis for quantum computation and their entangled and superimposed states are exploited to achieve quantum speed ups during computations. Various quantum circuits and gates  are created using qubits to achieve the quantum computation\cite{harel2007introduction}.  The power of quantum algorithms comes from taking advantage of quantum parallelism
and entanglement. 
 
The quantum gates are manipulated to carry the computation using quantum registers.   The most common quantum gates operate on spaces of one or two qubits, just like the common classical logic gates operate on one or two bits.  Operation of these quantum gates is what makes the quantum computing possible.

\begin{figure}{}
\centering 
\includegraphics[width=5cm]{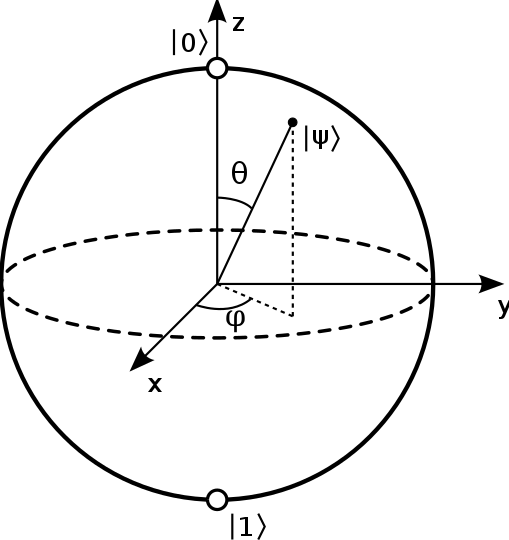}
\captionof{figure}{Bloch sphere.}
\label{Bloch-sphere}
\end{figure}

 The state of a quantum bit is described by a complex unit vector $| a \rangle$ in a $2$-dimensional Hilbert space.  The vector representation of a single qubit is $|\psi \rangle =  \alpha | 0 \rangle + \beta |1 \rangle$, where $\alpha, \beta \in \mathbb C$ and $| 0 \rangle$, $ |1 \rangle$ in the two dimensional Hilbert space $\mathcal H^2$.   Every 1-qubit pure state is represented as a point on the surface of the Bloch sphere, or equivalently as a unit vector whose origin is fixed at the center of the Bloch sphere, see Fig.~\ref{Bloch-sphere}. 


%

Some of the most used $1$-cubit gates are: the NOT gate (sometimes called Pauli X gate), Pauli gates which  are rotation gates and correspond to rotations about the $x$, $y$, and $z$-axes of the Bloch sphere, Hadamard gate, and the $\frac{\pi}{8}$-phase T gate, see \cite{nielsen2010quantum} for more details.   An important step in quantum algorithms is to use Hadamard gates to create superposition states. 

The vector representation of two qubits is $|a b  \rangle =  |a \rangle  \otimes |b \rangle  = \alpha_{00} | 0 0 \rangle + \alpha_{01}| 01 \rangle + \alpha_{10}| 10 \rangle + \alpha_{11} |11\rangle$, where $\alpha_{ij}  \in \mathbb C$.  The CNOT gate is one of the most important gates in quantum computing and it operates on quantum registers  consisting of 2 qubits. The spacial feature of the CNOT gate is that it flips the second qubit if only and only if the first bit is 1\cite{divincenzo1995two}.  
That is, the CNOT gate transform is described via the map 
\[
a |00 \rangle + 
b |01 \rangle + 
c |10 \rangle + 
d |11 \rangle  \mapsto 
a |00 \rangle + 
b |01 \rangle + 
c |11 \rangle + 
d |10 \rangle, 
\] 
which can also be represented by the permutation matrix 
\[ 
\text{CNOT} = 
\begin{pmatrix}
1 & 0 & 0& 0 \\ 
0 & 1 & 0& 0 \\ 
0 & 0 & 0& 1 \\ 
0 & 0 & 1& 0 \\
\end{pmatrix}. 
\] 
Operation of these quantum gates is what makes the quantum computing possible.

In the following figure we are giving a quantum circuit  using Hadamard gates to generate computational states of $3$ qubits.  As we mentioned above,  Hadamard gate is a single qubit transformation  and is given as follows
%
%
\[ \begin{split} 
H: & |0 \rangle \mapsto | + \rangle = \frac{1}{\sqrt 2}   \left( \frac{}{} | 0 \rangle  + |1 \rangle \right),  \\
& |1 \rangle \mapsto | - \rangle = \frac{1}{\sqrt 2}   \left( \frac{}{} | 0 \rangle  - |1 \rangle \right), 
\end{split}\]
which produces a superposition of $|0 \rangle$ and $|1 \rangle$.

\newcommand{\ket}[1]{\ensuremath{\left|#1\right\rangle}} 
\begin{figure}[h!]
	\centerline{
		\begin{tikzpicture}[thick]
		\tikzstyle{operator} = [draw,fill=white,minimum size=1.5em] 
		\tikzstyle{phase} = [fill,shape=circle,minimum size=5pt,inner sep=0pt]
		\tikzstyle{surround} = [fill=blue!10,thick,draw=black,rounded corners=2mm]
		%
		\node at (0,0) (q1) {\ket{0}};
		\node at (0,-1) (q2) {\ket{0}};
		\node at (0,-2) (q3) {\ket{0}};
		%
		\node[operator] (op11) at (1,0) {H} edge [-] (q1);
		\node[operator] (op21) at (1,-1) {H} edge [-] (q2);
		\node[operator] (op31) at (1,-2) {H} edge [-] (q3);
		%
		\node[phase] (phase11) at (2,0) {} edge [-] (op11);
		\node[phase] (phase12) at (2,-1) {} edge [-] (op21);
		\draw[-] (phase11) -- (phase12);
		%
		\node[phase] (phase21) at (3,0) {} edge [-] (phase11);
		\node[phase] (phase23) at (3,-2) {} edge [-] (op31);
		\draw[-] (phase21) -- (phase23);
		%
		\node[operator] (op24) at (4,-1) {H} edge [-] (phase12);
		\node[operator] (op34) at (4,-2) {H} edge [-] (phase23);
		%
		\node (end1) at (5,0) {} edge [-] (phase21);
		\node (end2) at (5,-1) {} edge [-] (op24);
		\node (end3) at (5,-2) {} edge [-] (op34);
		%
		\draw[decorate,decoration={brace},thick] (5,0.2) to
		node[midway,right] (bracket) {$\frac{\ket{000}+\ket{111}}{\sqrt{2}}$}
		(5,-2.2);
		\end{tikzpicture}
	}
	\caption{
		A quantum circuit  using Hadamard gates to generate computational states}
\end{figure}
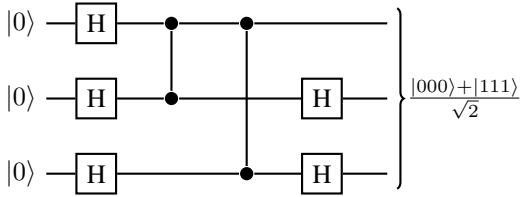

When quantum operations are performed in parallel we compute the tensor product.
In Fig. $4$, we apply Hadamard transform to the tensor product of three inputs, but since there is another Hadamard transform for the second and the third tensor components, we apply Hadamard transform again, obtaining:  
\begin{align*}
|0 \rangle &\mapsto \frac{|0 \rangle + |1\rangle }{\sqrt{2}} 
				\mapsto \frac{|0 \rangle + |1\rangle }{\sqrt{2}},   \\ 
|0 \rangle &\mapsto \frac{|0 \rangle + |1\rangle }{\sqrt{2}} 
				\mapsto |0\rangle,   \\ 
|0 \rangle &\mapsto \frac{|0 \rangle + |1\rangle }{\sqrt{2}} 
				\mapsto |0\rangle.   \\  
\end{align*}
This gives us 
\[ 
\frac{1}{\sqrt{2}}(|000\rangle + |100 \rangle).  
\] 
Now, since there is a CNOT gate for the first and the second tensor components, if there is a $0$ in the latter tensor (with a $1$ in the former tensor), we swap that to $1$. 
This gives us 
\[ 
\frac{1}{\sqrt{2}}(|000\rangle + |110 \rangle).  
\] 
Since there is a CNOT gate for the first and the third tensor components and since the latter component is a zero, we swap that to $1$. We thus obtain 
\[ 
\frac{1}{\sqrt{2}}(|000\rangle + |111 \rangle).  
\]

\section{Quantum parallelism}

Let $U_f$ be a quantum gate array that implements an arbitrary function $f$ with $m$ input and $k$ output bits which can be implemented on a quantum computer. If $U_f$ is applied to an input which is in superposition, then since $U_f$ is a linear transformation, it is applied to all basis vectors in the superposition simultaneously and will generate a superposition of the results given as follows
\[ \begin{split}
U_f  |x, y \rangle &\to |x, y \oplus f(x) \rangle \\
\sum_x a_x |x, 0 \rangle & \to \sum_x a_x |x, f(x)\rangle
\end{split}   \]
  In this way, it is possible to compute $f(x)$ for $n$ values of $x$ in a single application of $U_f$. 
  That is, we input many evaluations of $f$ in one unit time, which is equivalent to evaluating a function on a random input value. 
  We cannot obtain all output values since quantum states cannot be measured completely. 
  We can however obtain quantities that depend on many outputs $f(x)$. 
  This effect is
called quantum parallelism.

For example, consider the constant functions
\[ f(x) = 0, \, \, f(x) = 1.  \]
This is called a balanced function since the outputs are opposite for half the inputs. Since, $x$ is a qubit it can be in a superposition state, which as we mentioned in the previous section can be obtained by the application of the Hadamard gate. Let us start with a $|0 \rangle$ qubit and first apply the Hadamard gate and then apply the function $f$. 

\[
\begin{split}
 U_f  H |0 \rangle  &=   U_f\left(  \frac{  | 0 \rangle  + |1 \rangle }{\sqrt 2}   \right) 
 \hspace{3mm} \text{apply Hadamard}  \\
 & =  \frac{  | 0, 0 \oplus f(0)  \rangle  + |1, 0 \oplus f(1) \rangle }{\sqrt 2} 
 \hspace{3mm} \text{apply $U_f$}  \\
  &=   \frac{  | 0, f(0)  \rangle  + |1, f(1) \rangle }{\sqrt 2} 
  \hspace{3mm} \text{superposition of all } (x, f (x) ) \\
  & =  \frac{  | 0 0  \rangle  + |1 1\rangle }{\sqrt 2}
\end{split}
\]

We can use quantum parallelism and interference to obtain information that depends on many output values. 

%
%
%

\section{Quantum optimization models for machine learning}

\subsection{Quantum machine learning}

The performance of machine learning algorithms, often dealing with high dimensional vector spaces, suffer in performance using classical computers. Quantum parallelization and quantum associative memories have proven to accelerate the computations and can be deployed to solve complex machine learning algorithms both for optimization and learning\cite{purushothaman1997quantum,ezhov2000quantum}.  It is well known that the quantum computers are good at handling tensor and dot products in higher dimensions, and this feature will help the machine learning algorithms to compute using less computation time \cite{schuld2015introduction}. Shor has proved that the quantum computers can factor integers in polynomial time \cite{jozsa2003role}, which normally is not possible by classical computers. In machine learning, many optimization problems are intractable and cannot be computed in a given time and by a given computational resource. Heuristic optimization techniques are applied to solve such intractable complex problems. However, quantum computers can speed up the computation of such problems.

\subsection{Mathematical models for quantum machine learning} 

In this section, we describe how quantum computation can be used for higher-dimensional data sets. 
We then give an application towards image processing as an example. 
We also propose quantum parallelism with three possible decisions that one can make with the set of output vectors. 

Given a higher-dimensional data set encoded using a vector 
$\vec{f}^{(s)} =(v_1^{(s)},\ldots, v_P^{(s)})^t$ for each time $s$, 
where $s_0 \leq s \leq s_T$, 
we apply quantum transformations such as Fourier, Hadamard, and Haar wavelet transforms, as well as apply other operations needed to obtain a collection of output vectors simultaneously via the technique of quantum parallelism. 
What we do with the collection of output vectors depends on the original higher-dimensional data set. We may take the averages of the components, or we may look at the maximum or the minimum values that appear in each component, thus creating and working with a new output vector.

Next, we give an example of quantum techniques applied to an image representation. 
Given a $3$-dimensional image $F=(v_{ijk})_{M\times L\times N}$ 
encoded in a three-dimensional matrix at time $s$, where $s_0 \leq s\leq s_T$, 
where $v_{ijk}$ represents the pixel value at position $(i,j,k)$ with 
$1\leq i\leq M$, $1\leq j\leq L$, and $1\leq k\leq N$, 
a vector $\vec{f}^{(s)}$ time-stamped at $s$ with 
$MLN$ components is defined to be 
\begin{equation}
\begin{split}
\vec{f}^{(s)} 
&= (v_{111}^{(s)}, v_{211}^{(s)}, \ldots, v_{M11}^{(s)}, v_{121}^{(s)},
\ldots, \\ 
&\qquad  \quad 
v_{ij1}^{(s)},\ldots, v_{ML1}^{(s)},\ldots, v_{MLN}^{(s)})^t, 
\end{split}
\end{equation}
where we fix the right-most coordinate and read the $3$-dimensional matrix as a $2$-dimensional matrix by going down each column from left to right, 
and $s_0 \leq s \leq s_T$.  
The image data $\vec{f}^{(s)}$ 
may be mapped onto a pure quantum state 
$| f^{(s)} \rangle = \sum_{}^{} c_k^{(s)} |k\rangle$, 
where the computational basis 
$| k \rangle$ encodes the position $(i,j,k)$ of each pixel, 
and the coefficient $c_k^{(s)}$ encodes the pixel value, 
i.e., $c_k^{(s)}=v_{ijk}^{(s)}/(\sum_{i,j,k} (v_{ijk}^{(s)})^2)^{1/2}$ 
for $k< MLN$ and $c_k^{(s)}=0$ otherwise. 
We normalize the coefficients so that the resulting vector is in a quantum state. 
 
Now, we discuss quantum image transforms that may be applied to a vector in a quantum state.  Let $M=2^m$ and $L=2^l$, and $N=2^n$ with $\nu=m+l+n$. 
So the image has $2^{\nu}$ pixels. 
Image processing on a quantum computer corresponds to evolving the quantum state $|f^{(s)}\rangle$. 
Since a large class of image operations is linear in nature, 
the linear transformation in the quantum context is represented as 
$|g^{(s)}\rangle = U^{(s)}|f^{(s)}\rangle$, 
where $|f^{(s)}\rangle$ is the input image state and 
$|g^{(s)}\rangle$ is the output image state.

 In the final stage of quantum image processing, 
 we want to extract useful information from the output state. 
Furthermore, one is often not interested in $|g^{(s)}\rangle$ 
 itself but in some statistical characteristics or useful global features about the original data.

We now apply quantum edge detection algorithm, 
which is a pattern recognition task to recognize boundaries in an image, i.e., color intensity changes between two adjacent regions. 
Although classical algorithms require a computational complexity of at least 
 $\mathcal{O}(2^{\nu})$ since each pixel needs to be processed, 
 a quantum algorithm provides an exponential speed-up compared to existing edge extraction algorithms.

One algorithm that finds the boundaries between two regions in linear time,  
 independent of the image size, is as follows:  
 a Hadamard gate $H$, which is defined to be 
\[ 
H = \frac{1}{\sqrt{2}} 
\begin{pmatrix}
1 & 1 \\ 
1 & -1 \\ 
\end{pmatrix}, 
\]  
converts a qubit 
 $|0\rangle$ to $(|0\rangle + |1\rangle )/\sqrt{2}$ and 
 $|1\rangle$ to $(|0\rangle - |1\rangle )/\sqrt{2}$. 
Assume that the pixels in a black and white image have amplitude values 
$0$ and $1$, respectively, so 
each pixel is encoded in a binary form. 
 Since the positions of any pair of neighboring pixels in an image data are given by the sequences 
 $a_1\cdots a_{P-1}0$ 
 and  
$a_1\cdots a_{P-1}1$, with $a_i\in \{ 0,1\}$,  
their pixel values are stored as the coefficients 
$c_{a_1\cdots a_{P-1}0}^{(s)}$ and 
$c_{a_1\cdots a_{P-1}1}^{(s)}$  
of the corresponding computational basis states. 
The Hadamard transform changes these states to the new coefficients 
$c_{a_1\cdots a_{P-1}0}^{(s)} \pm c_{a_1\cdots a_{P-1}1}^{(s)}$. 

To elaborate in more detail, 
we first embed this vector into the higher-dimensional space $\mathbb{R}^{2P}$ by doubling all coordinates. 
We then apply the Hadamard transform $H$ tensored with the $P\times P$ identity matrix 
$H\otimes I_{P}$, where $I_{P}$ is the $P\times P$ identity matrix. 
Finally, we project onto the components that contain the differences of the coefficients to obtain the output state: 
\begin{equation*} 
\tiny 
\begin{pmatrix} 
c_0^{(s)} \\ 
c_1^{(s)} \\ 
c_2^{(s)} \\ 
c_3^{(s)} \\ 
\vdots \\ 
c_{P-2}^{(s)} \\ 
c_{P-1}^{(s)} \\ 
\end{pmatrix} 
\mapsto 
\begin{pmatrix} 
c_0^{(s)} \\ 
c_1^{(s)} \\ 
c_1^{(s)} \\
c_2^{(s)} \\ 
c_2^{(s)} \\ 
c_3^{(s)} \\ 
\vdots \\ 
c_{P-1}^{(s)} \\ 
c_{P-1}^{(s)} \\ 
c_0^{(s)} \\ 
\end{pmatrix} 
\mapsto 
\frac{1}{\sqrt{2}} 
\begin{pmatrix}
c_0^{(s)} + c_1^{(s)} \\   
c_0^{(s)} - c_1^{(s)} \\  
c_1^{(s)} + c_2^{(s)} \\ 
c_1^{(s)} - c_2^{(s)} \\ 
c_2^{(s)} + c_3^{(s)} \\  
c_2^{(s)} - c_3^{(s)} \\
\vdots \\ 
c_{P-1}^{(s)} + c_{0}^{(s)} \\ 
c_{P-1}^{(s)} - c_{0}^{(s)} \\  
\end{pmatrix}
\mapsto 
\frac{1}{\sqrt{2}} 
\begin{pmatrix}
c_0^{(s)} - c_1^{(s)} \\  
c_1^{(s)} - c_2^{(s)} \\ 
c_2^{(s)} - c_3^{(s)} \\
\vdots \\ 
c_{P-2}^{(s)} - c_{P-1}^{(s)} \\  
c_{P-1}^{(s)} - c_{0}^{(s)} \\  
\end{pmatrix}, 
\end{equation*}
which contains the full boundary information at time $s$. 
The final output state actually contains additional information, 
which is about the borders of the image, 
but we can project and remove such components from the final state. 
For a $\nu$-qubit input image state where $P=2^{\nu}$, 
$|f^{(s)}\rangle = \sum_{k=0}^{P-1} c_k^{(s)} |k\rangle$.

We are interested in the difference 
$c_{a_1\cdots a_{n-1}0}^{(s)} - c_{a_1\cdots a_{n-1}1}^{(s)}$ 
since if two pixels belong to the same region, then their difference vanishes; 
otherwise, their difference is nonvanishing, which indicates a boundary of a region. 
Thus the state of these qubits encodes the boundaries of an image at time $s$. 
This procedure yields a quantum state, 
encoding the information about the boundary.    
A measurement of a single local observable suffices since the goal is to discover if a specific pattern is present in the image.

Quantum parallelism makes it possible to do several parts of an algorithm simultaneously in order to complete a problem using less computation time. 
We now apply quantum parallelism on the output state described above, 
where we are interested in some property of all the inputs, 
not just a particular one.    
Thus we input the vectors of values for all time $s$ simultaneously, 
giving us a collection of vectors, which we will denote by 
\begin{equation*} 
\frac{1}{\sqrt{2}} 
\begin{pmatrix}
c_0^{(s_0)} - c_1^{(s_0)} \\  
c_1^{(s_0)} - c_2^{(s_0)} \\ 
c_2^{(s_0)} - c_3^{(s_0)} \\
\vdots \\ 
c_{P-1}^{(s_0)} - c_{0}^{(s_0)} \\  
\end{pmatrix}, 
\ldots,  
\frac{1}{\sqrt{2}} 
\begin{pmatrix}
c_0^{(s_T)} - c_1^{(s_T)} \\  
c_1^{(s_T)} - c_2^{(s_T)} \\ 
c_2^{(s_T)} - c_3^{(s_T)} \\
\vdots \\ 
c_{P-1}^{(s_T)} - c_{0}^{(s_T)} \\  
\end{pmatrix}. 
\end{equation*} 

There are several options on what we can do with this collection of vectors. 
One option is to average out the components, 
thus obtaining an average of where the boundaries exist. 
This method is feasible if the image at a fixed time is not sharp and if $s_0 \approx s_T$. 
A second option is to compute a new vector, 
replacing each entry with the value that appears the maximum or the minimum number of times. 
There may be other possibilities, but we leave it as an open problem.

\section{Conclusion}

We have provided  a mathematical model for a quantum machine learning optimization problem that are too complex to be computed using classical computers. 
The model is to apply quantum algorithms to a higher-dimensional data sets, and apply quantum parallelism in order to complete the algorithm in less computation time. 

We illustrated this algorithm by giving an example,   
where we estimate and identify the boundary of a  
$3$-dimensional moving object using quantum techniques.

%
%
\bibliographystyle{IEEEtran}
\bibliography{qml}

\end{document}